\documentstyle[12pt,twoside,fleqn,epsfig,espcrc1]{article}

\def\y0{y^{(0)}}

\newcommand \beq{\begin{eqnarray}}
\newcommand \eeq{\end{eqnarray}}

\newcommand{\mnote}[1]{\marginpar{\tiny {}}}   %       Margin note

\begin{document}

\title{\bf Dynamics of Ultra-Relativistic Nuclear Collisions with Heavy Beams:
An Experimental Overview
}
\author{
   Peter Braun-Munzinger\\
   Gesellschaft f{\"u}r Schwerionenforschung\\
   64220 Darmstadt, Germany\\
and\\
   Johanna Stachel\\
   Physikalisches Institut\\
   Universit{\"a}t Heidelberg\\
   69120 Heidelberg, Germany\\
   }
%\date{\today}
\maketitle

\begin{abstract}

    \noindent We review, from an experimental point of view,  the
current status of ultra-relativistic nuclear collisions with heavy beams.

\end{abstract}

%\vfil\eject

\section{Introduction} 

Reactions between heavy nuclei at ultra-relativistic energies have now
been studied for a number of years at the BNL AGS (11.4 GeV/nucleon) and
at the CERN SPS (158 GeV/nucleon). Progress has been rapid but it is
only in the last year that sufficiently complete and comprehensive data
sets have become available that a "picture" is beginning to emerge. In
the following we will briefly review the experimental highlights since
the last Quark Matter conference at Heidelberg.  Particular emphasis
will be placed on the identification of possible collective behavior and
hydrodynamic flow as well as on the question whether or not there is
local equilibrium at freeze-out. As most progress since QM'96 in the
leptonic sector will only
be reported at this conference  we will only touch upon recent
experimental developments in the study of the low-mass dilepton
continuum and very briefly summarize the status of anomalous charmonium
suppression. 

\section{Global Variables}

\subsection{ E$_T$ and dE$_T$/d$\eta$ Distributions}
Measurements of the transverse energy production and its spatial
distribution  in central  Au+Au collisions at the 
AGS \cite{e8771} and Pb+Pb collisions at the SPS \cite{na491} have
yielded maximum pseudorapidity densities of dE$_T$/d$\eta =$ 200 and 450
GeV, respectively. Using a 
simple Bjorken-type estimate these values imply energy densities in the
fireball formed in the collision of about 1.3 and 3 GeV/fm$^3$. Recent
results from solutions of QCD on the lattice \cite{laermann} imply
critical temperatures (for systems including dynamical quarks) well
below 200 MeV. The corresponding critical energy density is then of the
order of 1-1.5 GeV/fm$^3$. The fireball's parameters discussed above are
obviously in an interesting region. 

\subsection{N$_c$ and dN$_c$/d$\eta$ Distributions}
Pseudo-rapidity distributions of charged particles have been measured in
Au+Au \cite{e8772} and Pb+Pb \cite{na492} collisions over nearly the
full solid angle. The total charged particle multiplicity in central
collisions increases from about 450 at AGS energy to about 1500 at SPS
energy. At AGS energy there are about equal numbers of pions and
nucleons, while the pion to nucleon ratio is about 6 to 1 at SPS
energy. This implies (within a thermal model, see below) a 
significant increase of the entropy per baryon in the fireball from about 15
to 38.  

\section{Spectral Distributions}
Spectral distributions of high quality for protons and produced
particles now exist \cite{na492,e8661,e8773,e8911,wa981,na441} for heavy
colliding systems both at AGS and SPS energy. In the following we will
concentrate on the information on can glean from such spectra concerning
stopping and, in particular, the presence or absence of collective
features such as hydrodynamic flow. 
%As will be shown, the data exhibit
%indeed many features characteristic of collective flow. However, in a
%recent paper \cite{leonidov} an alternative interpretation of these
%features in the framework of an initial state scattering or random-walk
%picture was developed. We will confront these two pictures with the aim
%to find decisive differences.

\subsection{Rapidity Distributions and Baryon Stopping}
The rapidity distribution of identified baryons has been measured over
the full solid angle at AGS energy \cite{e8661,e8773}. The distribution
is approximately Gaussian in shape with a peak at central rapidity and a
width significantly narrower than that observed for the system Si+Al,
implying strong baryon stopping. A similar difference is observed for
the net proton rapidity distributions in Pb+Pb and S+S collisions at SPS
energy \cite{na492}. To get a complete picture here one has to await
measurements of the $\Lambda$ and $\bar \Lambda$ rapidity distributions.
Current results from Na49 \cite{roland} are still somewhat
controversial, with the $\Lambda$ distribution nearly as narrow in
rapidity as the $\bar \Lambda$ distribution. In any case, there is
significantly increased stopping also at SPS energies, which is, of
course, reflected in the increased entropy per baryon (see above). 

\subsection{Transverse Momentum Distributions}
A general feature which has emerged from many measurements of transverse
momentum distributions is that, especially for central collisions of
heavy nuclei, the invariant distributions $d^2N/m_tdm_tdy$ which are
approximately exponential in shape, i.e. $d^2N/m_tdm_tdy \propto
\exp(-m_t/T)$, have inverse slope constants $T$ increasing linearly with
the mass of the particle under consideration. Since the transverse
momentum $p_t = m \gamma\beta_t$ this fact has been widely interpreted
as evidence for collective transverse flow: if there is a common flow
velocity (or velocity profile) $\beta_{t}$ superimposed on the random
(thermal) motion of particles, the slope constant $T$, which is
proportional to $\langle p_t \rangle$ increases linearly with $m$. One
can, however, also reproduce many features of the measured spectra by
assuming that the incoming nucleons undergo initial state scattering
\cite{leonidov}. Since the amount of initial state scattering only
depends on the number of nucleon-nucleon collisions each nucleon
undergoes during the collision, hence is geometrical in nature, one does
not need to invoke any collective expansion to explain the data in this
approach. To resolve this issue, we discuss in some detail the two
models and their implications.

\subsubsection{Random Walk}

The aim of the random walk model \cite{leonidov} is to provide a
description of minimum bias data on transverse momentum distributions in
proton-nucleus collisions by a superposition of the "kicks" the
projectile nucleon 
undergoes in each nucleon-nucleon collision in the target nucleus. Since
the extrapolation to AA collisions is then entirely geometric, one can
test the predictions of this model directly by comparison to measured
transverse momentum distributions. For pA collisions the approach is as
follows. Each projectile nucleon undergoes $N_A$ nucleon-nucleon
collisions in the target. Since, averaged over impact parameter $b$, the
nuclear thickness of the target is 4/3$R_A$, where $R_A =1.12A^{1/3}$fm is
the radius of the target, 
\beq
 N_A = 4/3 R_A \sigma_{NN} n_0.
\label{walk1}
\eeq
Here $n_0= 0.17/{\rm fm^3}$ is the nuclear density and the total
nucleon-nucleon cross section 
$\sigma_{NN}$ is about 40 mb. One further assumes that each NN collision
produces a fireball and that all final particles are emitted from the
sequence of fireballs so created. Each fireball moves at a certain
rapidity $y$ and transverse momentum $p_t$, determined by a boost
invariant longitudinal expansion scenario (for $y$) and by the history of
NN "kicks" for $p_t$. To obtain the observed scaling with mass of the
inverse slope constants the random walk proponents assume that the
"kick" happens in transverse rapidity $\rho_t = 0.5
\ln((m_t+p_t)/(m_t-p_t))$ rather than in $p_t$. Although this physically
seems not obvious, it clearly introduces a "transverse velocity"
distribution and, 
consequently, a mass dependent slope parameter. Since the sequence of
kicks is stochastic, the distribution of the fireballs in transverse
rapidity is assumed to be
\beq
f_{pA}(\rho_t) \propto \exp(-\rho_t^2/\delta_{pA}^2),
\label{walk2}
\eeq

with 
\beq
\delta_{pA}^2=(N_A-1) \delta_0^2.
\label{walk3}
\eeq

Here, $\delta_0$ is the only free parameter, apart from the fireball
temperature (see below). It is determined by a fit to pA data and then
kept constant for AB collisions. 

Generalizing the above expressions to nucleus-nucleus collisions we
replace $\delta_{pA}^2$ by
\beq
\delta_{AB}^2(b) = (N_c(b)/A + N_c(b)/B -2) \delta_0^2
\label{walk4}
\eeq
with
\beq
N_c(b)=T_{AB}(b) \sigma_{NN}.
\label{walk5}
\eeq
Here, the total number of collisions $N_c(b)$ for a given impact
parameter $b$ is evaluated with the help of the nuclear thickness
function \cite{charm1}. The purely geometrical nature of the problem
becomes therefore transparent. Apart from a normalization factor which
does not concern us here the expression for transverse mass spectra then
reads:
\beq
\frac{dN^{AB}} {m_tdm_t} \propto \int d\rho_t \exp(-\rho_t^2/
\delta_{AB}^2) g(m_t),
\label{walk6}
\eeq
where 
\beq
g(m_t)=  \int_{-y_L}^{y_L} dy m_t cosh(y-y_{m}) I_0(\frac{p_t
sinh(\rho_t)} {T}) 
K_1( \frac{m_t cosh(y-y_{m})} {T} ).
\label{walk7}
\eeq
Here, $I_0$ and $K_1$ are the modified Bessel functions and the
particles are detected at rapidity $y_{m}$.

\begin{figure}[htb]
%\vspace{9cm}
% \special{psfile=./small_slopes.ps  hoffset=-10 voffset=-170 hscale=110 vscale=110 angle=0}      
%\vspace{-7.5cm}
\begin{center}
\epsfig{file=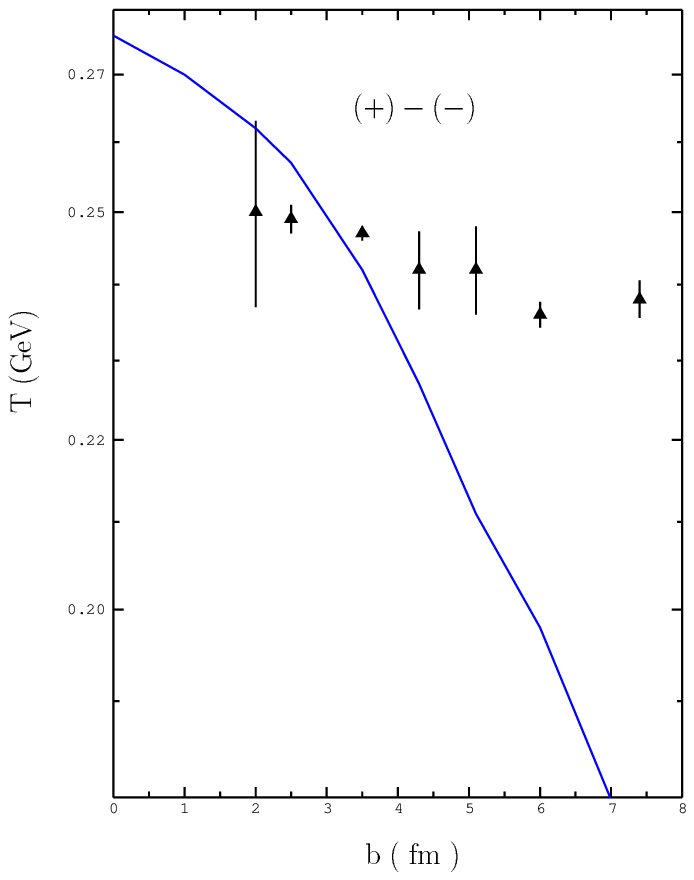,width=9cm}
\end{center}
 \caption{ }
Comparison of measured inverse slope constants and values calculated in
the random walk model for different centralities. Data are from the
CERES collaboration \cite{ceres1}.
 \label{fig:wk2}
\end{figure}

In the following, we will compare predictions according to
Eq. (\ref{walk6}) with 
recent data on "proton" transverse momentum distributions, {\it i.e.}
the difference of the spectral distributions of positively and negatively
charged particles, from the CERES
collaboration \cite{ceres1}.
\begin{figure}[tb]
%\vspace{9cm}
% \special{psfile=./small_spectra.ps  hoffset=10 voffset=-120 hscale=70 vscale=70 angle=0}      
%\vspace{1.5cm}
\begin{center}
\epsfig{file=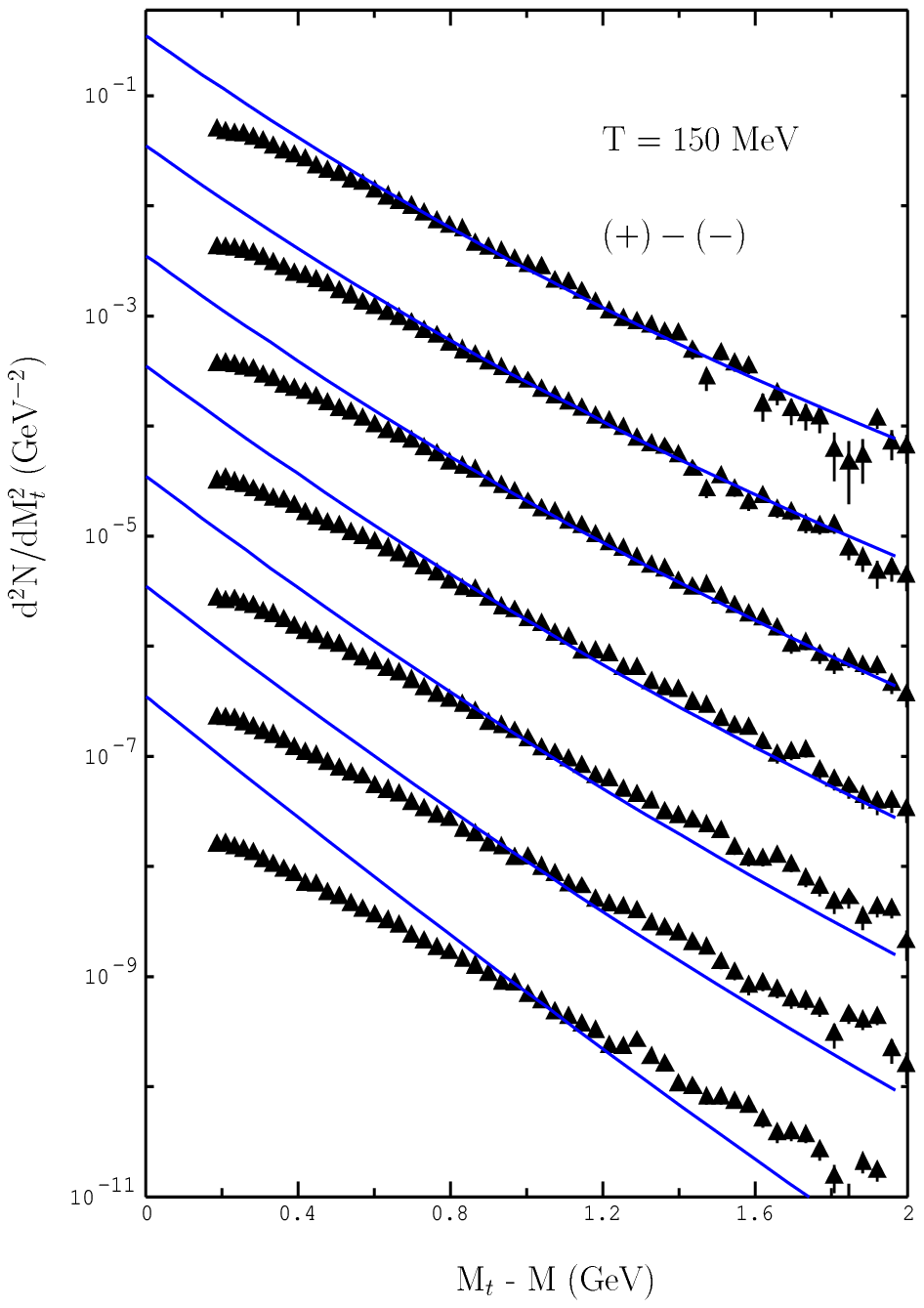,width=9cm}
\end{center}
 \caption{ }
Comparison of (+) - (-) spectra for Pb+Au collisions with predictions of
the random walk model for different centralities (increasing centrality
from bottom to top). The data are from the CERES collaboration \cite{ceres1}.
 \label{fig:wk1}
\end{figure}
These data for Pb+Au collisions were
recorded with a multiplicity trigger corresponding to the upper 35~\% of
the geometrical cross section. This multiplicity range
was subdivided into 7 exclusive intervals corresponding to (in a
geometrical interpretation) mean impact
parameters of 2, 2.5, 3.5, 4.3, 5.1, 6.0, and 7.4 fm, respectively. The
data are shown in Fig. \ref{fig:wk1}.Data in different multiplicity
(or impact 
parameter) bins were multiplied by consecutive factors of 8. The lines
in Fig. \ref{fig:wk1} correspond to predictions of the random walk model
according to Eq. (\ref{walk6}), with  $\delta_0 = 0.146$ and a fireball
temperature T = 150 MeV. Since we are
only interested in the shape of the distributions here, the
normalization of the calculation relative to the data is arbitrary. Note
that, for central 
collisions corresponding to mean impact parameters 3.5 fm and less, the
calculated distributions reproduce the shape of the data very well in
the range $0.7 < m_t -m < 2$ GeV. For more peripheral collisions,
however, the predictions by the random walk model exhibit inverse slope
constants which are considerably smaller than seen in the data. 
It is important to realize that there is no parameter to change here, as
the centrality dependence 
is determined by the thickness of matter traversed. If one determines
the inverse slope constants from a fit to both the data and the
calculations  in 
the range $0.6 < m_t -m < 2$ GeV  one can make this comparison more
quantitative, as shown in Fig. \ref{fig:wk2}.
Here, the discrepancy between data 
and calculations becomes very obvious: certainly in its present form the
random walk prescription is not consistent with the observed very weak impact
parameter dependence of inverse slope constants measured for protons in
Pb+Au collisions at SPS energy. Taking into account further that the
calculated spectra also deviate strongly from the data at lower $m_t -m$
values we conclude that the random walk model is not anymore a contender
to describe the flow-like features observed in the transverse momentum spectra.

\subsubsection{Transverse Flow}
\begin{figure}[htb]
\begin{center}
\epsfig{file=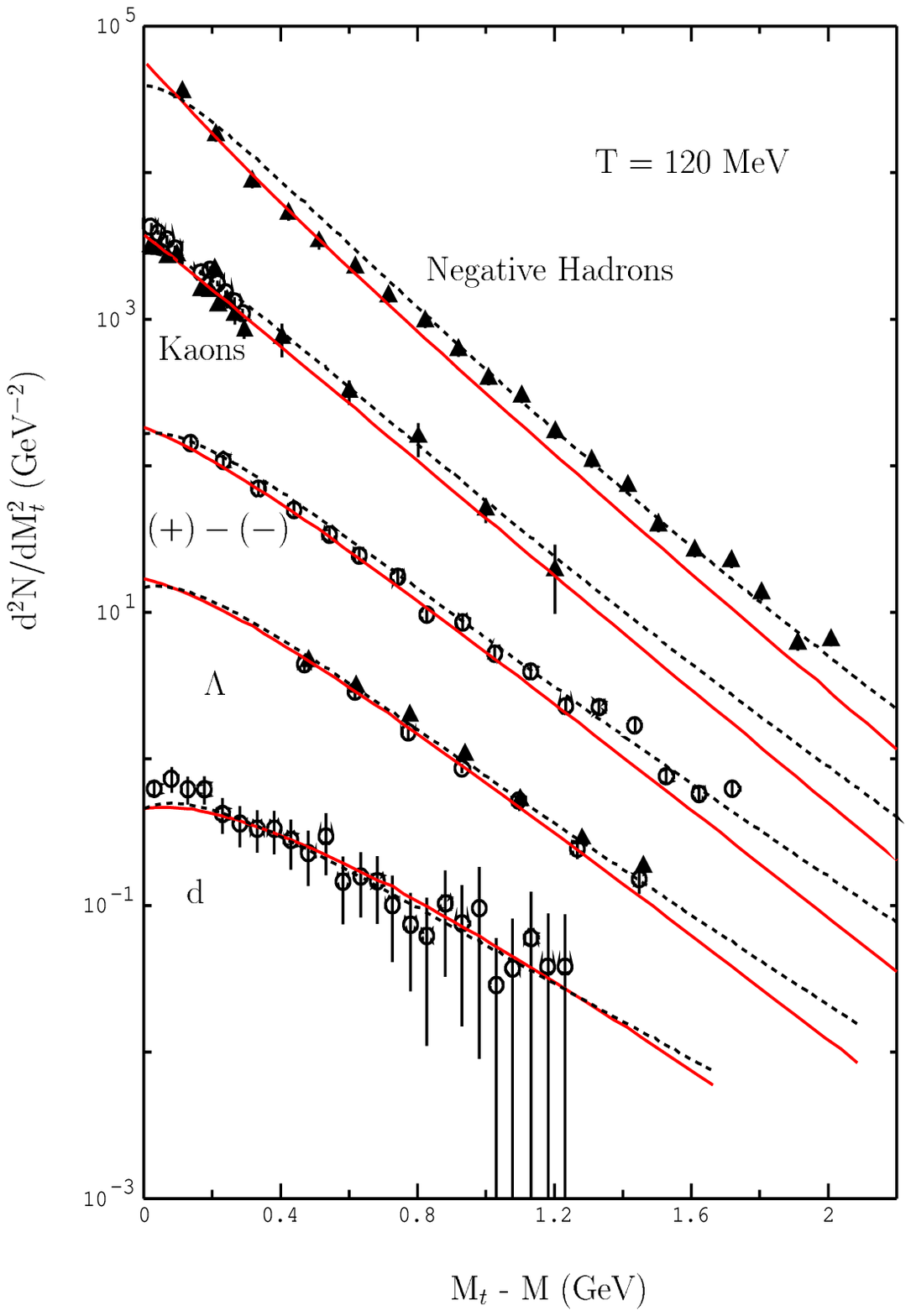, width=8.5cm}
\end{center}
\caption{ }
Comparison of hydrodynamic calculations assuming a freeze-out temperature
of 120 MeV with transverse momentum spectra
for various particles. The data are from the NA49 collaboration
\cite{roland} and are arbitrarily renormalized for clarity of
presentation. The solid lines are calculations assuming a linear
velocity profile. The dashed lines use the velocity profile determined
by the initial conditions and the equation of state. For details see text.
 \label{fig:trans1}
\end{figure}

\begin{figure}[htb]
\begin{center}
\epsfig{file=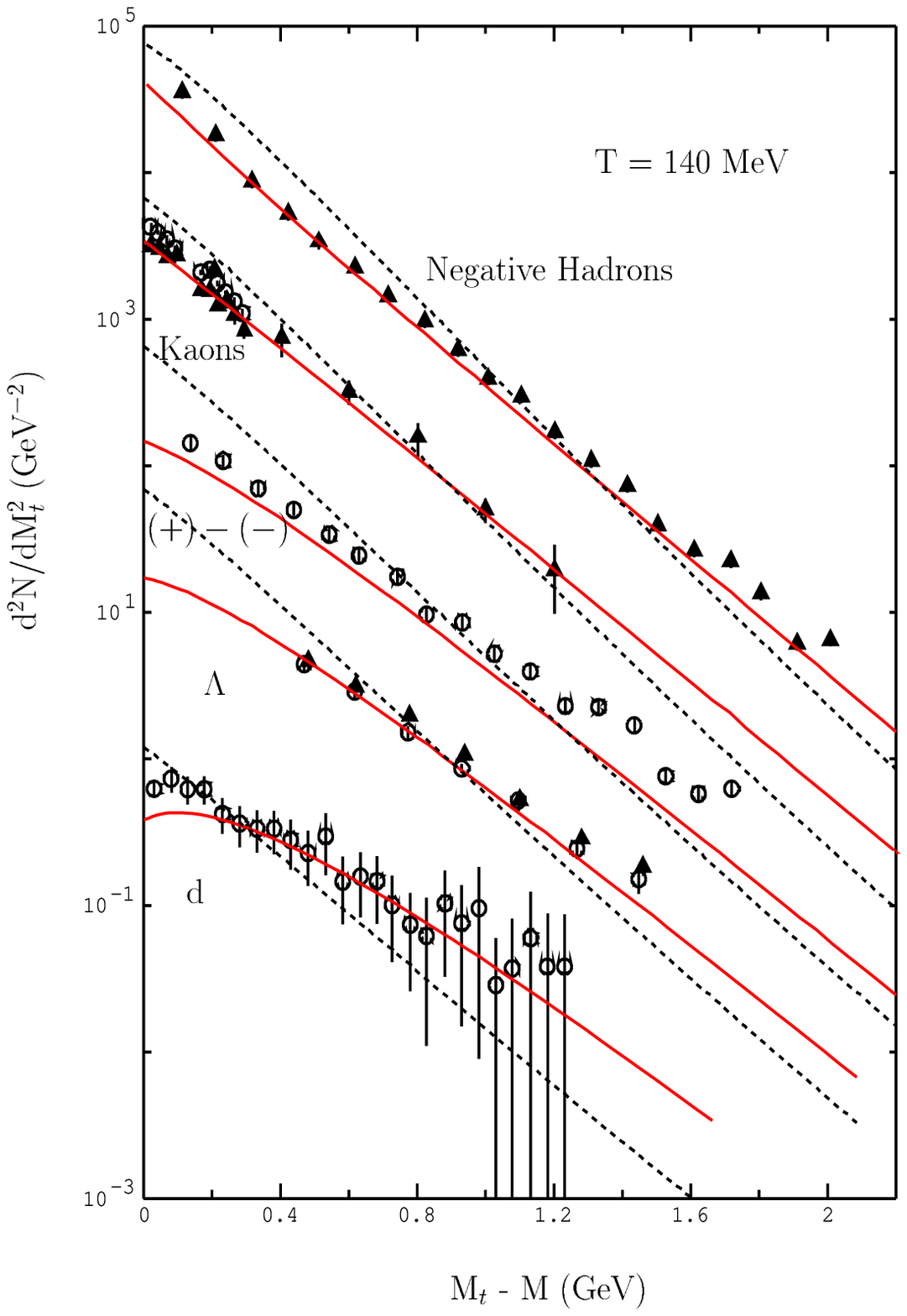, width=8.5cm}
\end{center}
 \caption{ }
Same as Figure 3 but for a freeze-out temperature of 140 MeV.
 \label{fig:trans2}
\end{figure}

For central collisions, the transverse momentum spectra of all observed
particles can be well described in a hydrodynamical approach. The basic
equations for azimuthally symmetric flow, i.e. central collisions, are
very similar to Eq. \ref{walk6}, except for the distributions in
transverse rapidity. In a hydrodynamical approach, these distributions
are obtained from the transverse flow velocity profile which, of course,
depends 
on the initial conditions as well as on the underlying equation of
state. We illustrate the success of this approach for data from the NA49
collaboration \cite{roland} in Figs. \ref{fig:trans1} and
\ref{fig:trans2}. The hydrodynamic 
calculations used the transverse velocity profile recently calculated by
Alam et al. \cite{cley1}. The initial conditions were chosen to
reflect the measured charged particle multiplicities in central Pb+Pb
collisions 
at SPS energy.
For the equation of state it was assumed that a pure
quark-gluon plasma undergoes a first order phase transition at $T_c$ =
160 MeV to a hadronic resonance gas which freezes out  at constant
temperature $T_f$.  
In Fig. \ref{fig:trans1} the data
are compared to the calculations represented by the dashed lines for
$T_f$ = 120 MeV. Excellent 
agreement between the shape of the measured distributions and the
calculations is obtained for all particle species ranging from pions to
deuterons. We also show, in Fig. \ref{fig:trans1}, the result of
calculations with a linear profile $\tanh{\rho_t} =\beta_t= \beta_{max}
\frac{r} {R_A}$. Again, good agreement between data and calculations is
obtained for $T_f$ = 120 MeV and $\beta_{max}= 0.6$. As is well known
$T_f$ cannot easily be determined from such fits, as higher $T_f$ values
can be traded off against lower $\beta_t$ values. This is illustrated in
Fig. \ref{fig:trans2} where the solid lines show the results of a
calculation with $T_f$ = 140 MeV and $\beta_t = 0.45$. We note, however,
that the full hydrodynamic calculation does not exhibit this degeneracy:
for $T_f$ = 140 MeV, there is, with the present initial conditions and
equation of state, no possibility to describe all measured distributions
simultaneously. Under the assumptions discussed above, the data seem to
clearly favor a relatively low freeze-out temperature near 120 MeV. The
reason is that, to build up the flow in the hadronic phase, one needs
the freeze-out to happen not too close to $T_c$. A
similar conclusion was recently obtained by the NA49 collaboration from
a simultaneous analysis of tranverse momentum spectra and two-pion
interferometry data \cite{appels}.

\subsection{Azimuthal Anisotropy and Directed Flow}

\begin{figure}[htb]
%\vspace{10cm}
% \special{psfile=./side1.ps hoffset=-20 voffset=-35 hscale=70 vscale=70 angle=0} 
%\vspace{-7.5cm}
\begin{center}
\epsfig{file=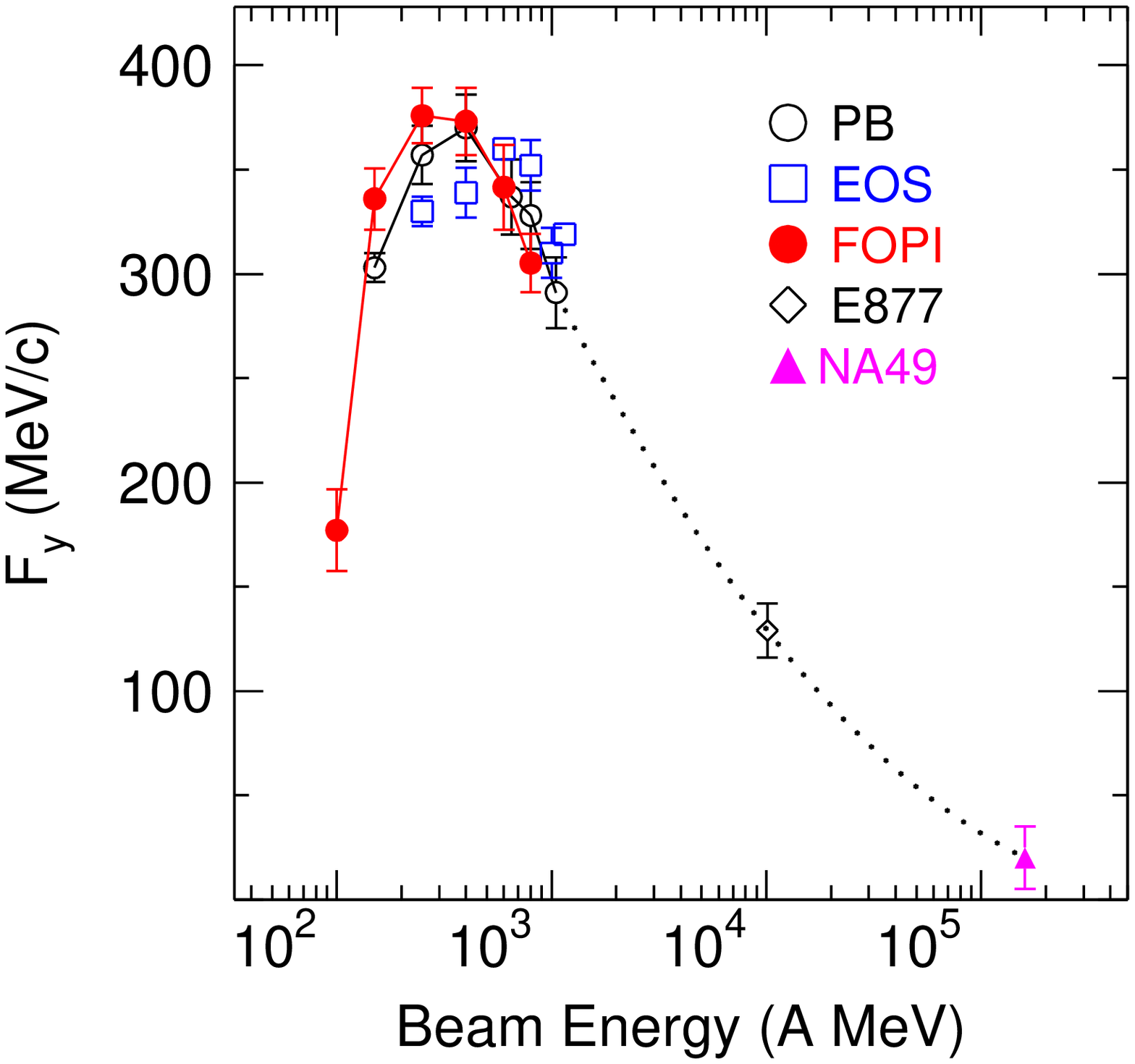, width=8.5cm}
\end{center}
 \caption{ }
Energy dependence from SIS to SPS energies of the proton sidewards flow
as a function of the beam kinetic energy per nucleon. 
 \label{fig:f1}
\end{figure}

The study of collective flow effects via the experimental observation of
anisotropies in azimuthal distributions for non-central collisions is
motivated by theoretical predictions that a quark-gluon phase will lead
to distinct differences in the flow patterns compared to what is
expected for  a hadron gas. Such phenomena were
originally studied experimentally at Bevalac energies \cite{reis} and
below. With 
the advent of the heaviest beams at the AGS and SPS such effects were
also established there \cite{flow877,flowna49}, at first in
distributions of global observables such at E$_T$ and later also in
distributions of identified particles. It is now customary to extract
information on flow by determination of the Fourier coefficients $v_i$ of the
azimuthal distributions $F(\phi)$ integrated over a certain rapidity interval
\beq
F(\phi) = F_0 (1 + \sum_{i=1}^{n} 2v_i \cos(i\phi)).
\label{flow1}
\eeq
The dipole coefficient $v_1$ is also denoted as 'directed flow' and is
associated to the mean transverse momentum in the reaction plane
$\langle p_x \rangle$ by
\beq
v_1 = \langle p_x \rangle/\langle p_t \rangle,
\eeq  
the quadrupole coefficient $v_2$ is also called 'elliptic flow'.
The odd Fourier coefficients change sign at midrapidity, while the even
coefficients are symmetric. 
A quantity that
was used to characterize the directed flow is the change 
of $\langle p_x \rangle$ with rapidity in the vicinity of mid-rapidity
and indeed the quantity 
\beq
F_y = d\langle p_x \rangle/dy
\eeq
is scale invariant and should be independent of beam energy if there is
no change in the physics. Fig. \ref{fig:f1} displays the information on
directed flow of protons in terms of this scale invariant variable as a
function of the beam kinetic energy per nucleon \cite{reis}. One can see
a plateau in 
the beam energy range of a few hundred MeV to 1 GeV and a steep fall-off
to much lower values at the AGS and again much lower values at the
SPS. It should be noted that the baryon density reached
in the collisions is maximal for the AGS regime while it is lower and of
comparable magnitude for the high end of the Bevalac/SIS regime and the
SPS. So, at a qualitative level, one can clearly state that there is a
difference in the equation of state over the energy regime considered
with an increasing softening at the higher energies.

It was shown at this conference \cite{flow877qm97} that the directed flow of
light nuclei is larger than the proton flow and increases monotonically
with particle mass. The pion and kaon flow on the other hand appear to
be a complex superposition of Coulomb effects, strong final state
interaction, and collective flow which is apparent in a distinctly
different $p_t$ dependence of $v_1$ (see \cite{qm97877,flow877qm97}).

At AGS energy it was shown that in a hadronic cascade code (RQMD) one
can indeed reproduce the overall magnitude of the proton directed flow
if a repulsive mean field between the nucleons is introduced
\cite{flow877}. The same 
calculation however fails completely to reproduce the shape of the $p_t$
dependence of the proton directed flow which is on the other hand well matched 
by a hydrodynamically inspired picture of a sideways moving and
expanding fireball \cite{flow877,qm97877}. \\
The elliptic flow of nucleons, quantified by the coefficient $v_2$,
shows an interesting dependence on beam energy. At lower beam energies
it reflects the energy dependence of the nucleon-nucleon interaction
which is attractive at very low energies and becomes  repulsive
in the 50 to 100 MeV/nucleon range. In terms of elliptic flow this is
reflected in a change of sign of $v_2$, i.e. a 90$^{\circ}$ change in
orientation of the ellipse (positive sign: long axis in the reaction
plane) and leads to the so-called squeeze-out in the Bevalac/SIS energy
regime with dominant nucleon emission perpendicular to the reaction
plane. At AGS energy the elliptic flow is small but now oriented along
the reaction plane \cite{flow877} and at SPS energy the elliptic flow is
significantly stronger but oriented the same way. Such behavior had been
predicted by theory \cite{sorge} as a consequence of the changing
relative importance of shadowing (leading to out-of-plane squeeze-out)
and of collective flow related to the pressure build-up early in the
reaction. At AGS and higher energies this compression related elliptic
flow appears to dominate. \nopagebreak The data are by now quantitative enough that a
theoretical analysis should be performed to deduce the early pressure 
build-up.
\begin{figure}[htb]
%\vspace{11cm}
% \special{psfile=./squeeze1.ps hoffset=0 voffset=0 hscale=70 vscale=70 angle=0} 
%\vspace{-1cm}
\begin{center}
\epsfig{file=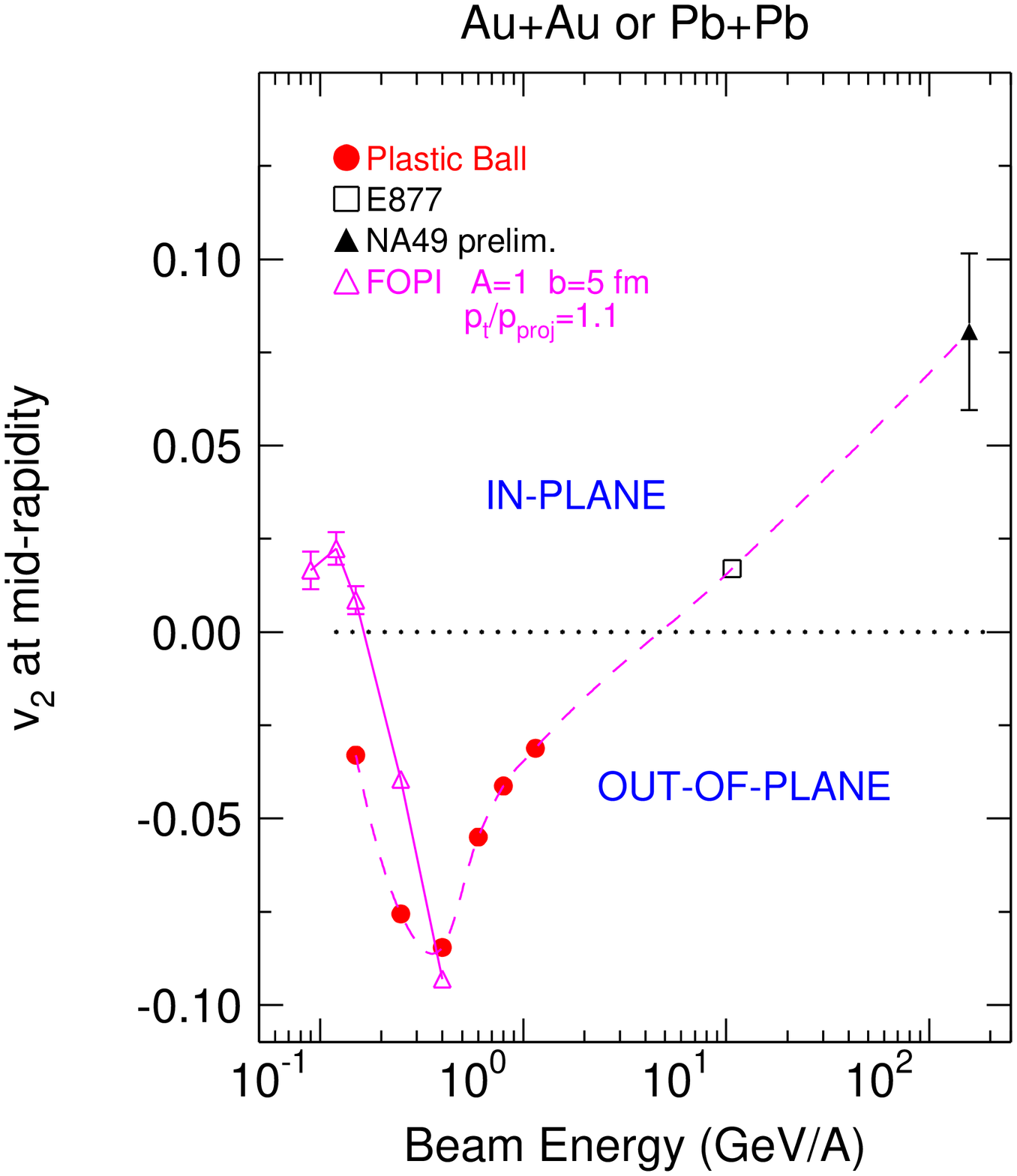, width=9cm}
\end{center}
 \caption{ }
Energy dependence from SIS to SPS energies of the elliptic 
flow (second Fourier coefficient). 
 \label{fig:f2}
\end{figure}

\subsection{Thermal Equilibrium and Freeze-out}

At the time of the Quark Matter '96 conference one of us reviewed the
evidence for chemical and thermal equilibrium among the hadrons produced
in ultra-relativistic collisions between heavy nuclei
\cite{stachel}. The conclusion at that time was that many features of
the data indeed imply that a large degree of chemical equilibration is
reached both at AGS and SPS energy. Since that time more data have become
available (essentially all at SPS energy) and we will provide a brief
update on the question of chemical equilibrium. We would like to
stress, however, that a final word on this important question can come
only after the data have been consolidated and when data cover as much
as possible of the full solid angle. 

In Table \ref{ratios} we show an update of presently available
particle ratios for Pb+Pb central collisions at SPS energy, along with
predictions from a thermal model calculation \cite{heppe} which is
slightly refined compared to our previous model \cite{bswx}. Fairly good
agrement between data and calculations is obtained assuming full
chemical (including strangeness) equilibration for temperatures of
160 and 175 MeV, corresponding baryon chemical potentials of  200 and
270 MeV, and no strangeness suppression. In fact, the more consolidated
data available now favor the larger baryon chemical potential; we mainly
show the calculation with T = 160 MeV and $\mu_b$ = 200 MeV as a
reference (which best described the preliminary data available at the
time of Quark Matter '96). The systematic uncertainties in
the particle ratios as witnessed by the differences, e.g., in the $\bar
\Lambda/\Lambda$ or $\Xi^-/\Lambda$ ratios
from NA49 and WA97 are presently still too large to make a final
judgement. We, therefore, feel that the claim of evidence for partial
strangeness equilibration  from a recent analysis of
NA49 data \cite{becat}  is premature. 

\begin{table}[h]
\caption{Particle ratios calculated in 2 versions of a thermal model for
temperatures of 160 and 175 MeV, and baryon chemical potentials $\mu_b$
of 200 and 270 MeV, in 
comparison to experimental data (with statistical errors in parentheses)
for central collisions of Pb+Pb at SPS energy. }
\vspace{0.5cm}
\label{ratios}
%\newpage
\begin{center}
\begin{tabular}{||c|ll|lcc||}
\hline
\hline
\multicolumn{1}{||c|}{Particles} & \multicolumn{2}{c|}{Thermal Model} &
\multicolumn{3}{c||}{Experimental Data}\\
 & 1 & 2 & exp. ratio & Exp. & y  \\ \hline
  p-\=p/neg$^a$ & 0.17 & 0.23 & 0.23(3) & NA49 & 0.2-5.6  \\
  $\pi^-/\pi^+$ & 1.04 & 1.06 & 1.10(5) & NA49 & all \\
  \=p/p$^a$ & 0.086 & 0.050 & 0.055(10) & NA44 & 2.3-2.9  \\
  \=p/p$^b$ & 0.099 & 0.076 & 0.085(8) & NA49 & 2.5-3.3  \\
  \=d/d & 6.7 $\cdot 10^{-3}$ & 2.1 $\cdot 10^{-3}$ & 3.6(8) $\cdot
10^{-3}$ & NA44 & 1.9-2.1\\ 
\hline
  K$^+$/K$^-$ & 1.57 & 1.94 & 1.61(15) & NA49 & 2.5-3.3  \\
   &  &  & 1.85(9) & NA44  & 2.4-3.5  \\
  K$^0_s/\pi^-$ & 0.145 & 0.138 & 0.125(19) & NA49 & all\\
  $\bar{\Lambda}/\Lambda^a$ & 0.147 & 0.115 & 0.128(12) & WA97 & 2.3-3.4 \\ 
  $\bar{\Lambda}/\Lambda^b$ & 0.17 & 0.15 & 0.19(1) & NA49 & 2.6-3.8 \\
  2$\phi/(\pi^++\pi^-)$ & 0.020 & 0.020 & 9.1(10) $\cdot 10^{-3}$ & NA49
& all \\  
\hline
  $\Xi^+/\Xi^{-a}$ & 0.255 & 0.273 & 0.266(28) & WA97 & 2.4-3.4 \\
  $\Xi^-/\Lambda^a$ & 0.114 & 0.101 & 0.093(7) & " & " \\ 
  $\Xi^-/\Lambda^b$ & 0.093 & 0.084 & 0.13(4) & NA49 & 2.0-2.6 \\
  $\Xi^+/\bar{\Lambda}^a$ & 0.198 & 0.239 & 0.195(23) & WA97 & 2.4-3.4 \\ 
\hline
  $\Omega^+/\Omega^{-a}$ & 0.46 & 0.68 & 0.46(15) & " & " \\
  $(\Omega^++\Omega^-)/(\Xi^++\Xi^-)^a$ & 0.154 & 0.168 & 0.195(28) & " & " \\
\hline
\hline
\end{tabular}
\end{center}

$^a$No feeding from weak decays.\\
%\vspace{-1.3cm}
$^b$ Feeding from weak decays included.

\end{table}

Note that, as discussed above, there is now evidence that, at SPS
energy, thermal freeze-out happens at smaller temperature than chemical
freeze-out.
\begin{figure}[htb]
\begin{center}
\epsfig{file=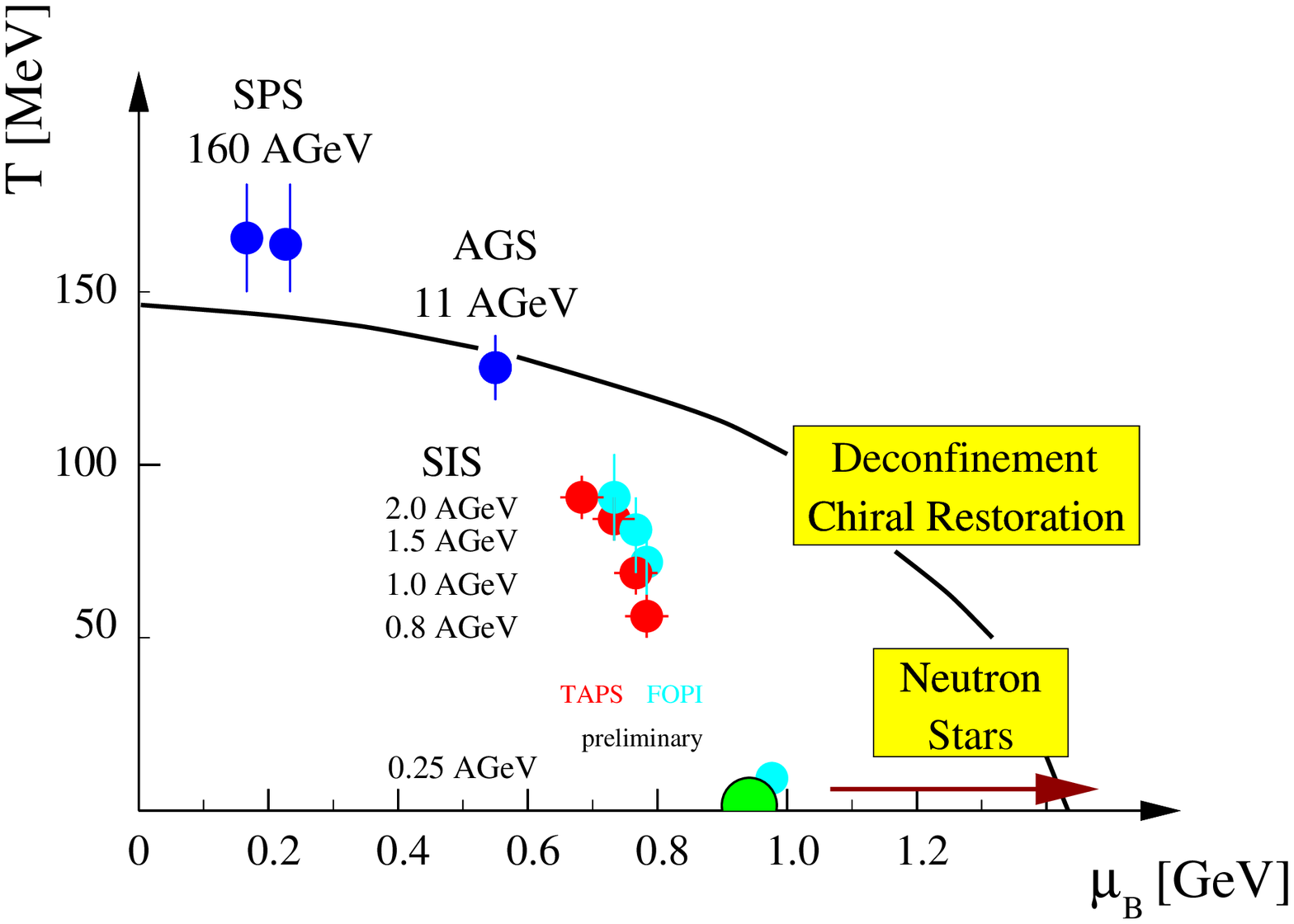, width=10.0cm}
\end{center}
 \caption{ }
Freeze-out parameters deduced from the hadro-chemical analysis of particle
production yields at SIS, AGS, and SPS energies. At the higher energies
the freeze-out points approach the phase boundary. 
 \label{fig:freeze-out}
\end{figure}

In Fig. \ref{fig:freeze-out} we show, for SIS, AGS, and SPS energies,
the freeze-out parameters deduced from the chemical analysis of particle
yields. It is interesting to note that, at the higher energies, the
freeze-out points approach the calculated phase boundary
\cite{bswx} while freeze-out parameters for experiments at SIS energies
never come close to it.

\section{Leptonic Observables}

\subsection{The dilepton continuum}

\begin{figure}[htb]
%\vspace{9cm}
% \special{psfile=./plb_4.ps  hoffset=30 voffset=20 hscale=50 vscale=50 angle=0} 
%\vspace{-1cm}
\begin{center}
\epsfig{file=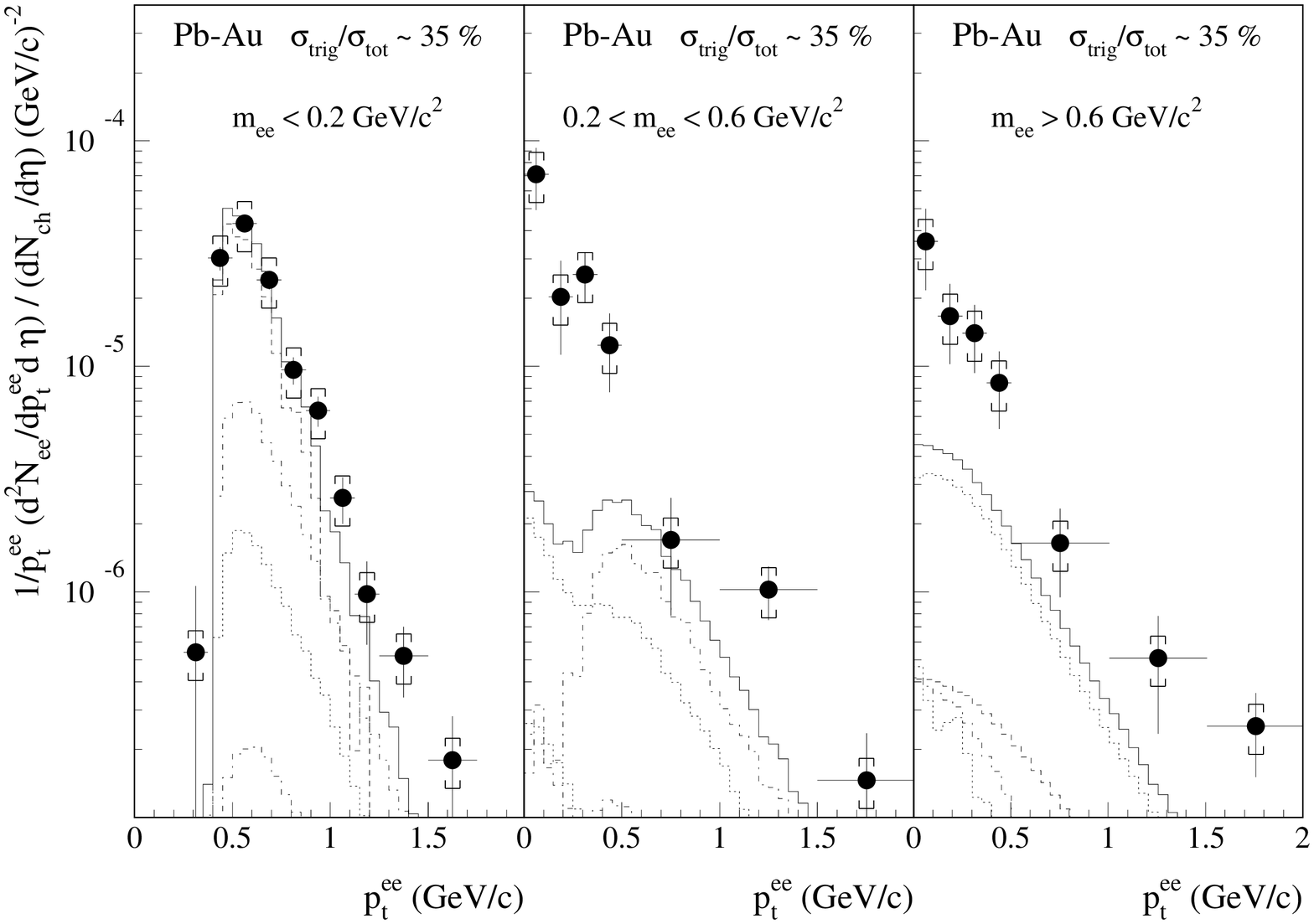, width=13.0cm}
\end{center}
 \caption{ }
The dependence on pair transverse momentum of the inclusive lepton-pair
continuum for different mass windows. The data are for Pb+Au collisions
from the CERES collaboration \cite{ceres3}.
 \label{fig:dilep1}
\end{figure}

First results from a measurement of the dilepton continuum at low mass in
Pb+Au collisions were shown at the Quark Matter '96 conference
\cite{ceres2}. Meanwhile, the CERES collaboration has released  the
final data \cite{ceres3}. Similar to what was observed for the S+Au
data, the electron pair yield in the invariant mass range $ 0.2 -
2.0$ GeV is enhanced by a factor of $3.5 \pm 0.4(stat.) \pm 0.9 (syst.)$
over what is expected from neutral meson decays extrapolating from
nucleon-nucleon collisions. The enhancement 
increases strongly (approximately quadratically) with charged particle
multiplicity and, hence, centrality. Most notable is that the observed
enhancement seems to be concentrated at low transverse momenta. This is
shown in Fig. \ref{fig:dilep1} where, for Pb+Au collisions, inclusive
e$^+$e$^-$ pair transverse momentum spectra are presented for three
different pair mass ranges. The spectra are normalized again to charged
particle multiplicity. For pair masses less than 0.2 GeV the data 
agree, as expected,  with the sum of the hadron decay contributions. For
larger pair masses, especially visible in the mass range between 0.2 and
0.6 GeV, the enhancement is strongest at very low pair transverse
momentum. 

These results corroborate  earlier findings for S+Au
collisions  of a lepton pair yield which is significantly enhanced
compared to expectations for neutral meson decays. The new results,
i.e. the strong centrality dependence and the concentration of the
enhancement at low pair transverse momenta, should help to distinguish
between various models put forward for the interpretation of these data. 

\subsection{Charmonium suppression}

\begin{figure}[htb]
%\vspace{14cm}
% \special{psfile=./jpsi_l.ps   
%          hoffset=20 voffset=-45 hscale=70 vscale=70 angle=0}
%\vspace{-1.5cm}
\begin{center}
\epsfig{file=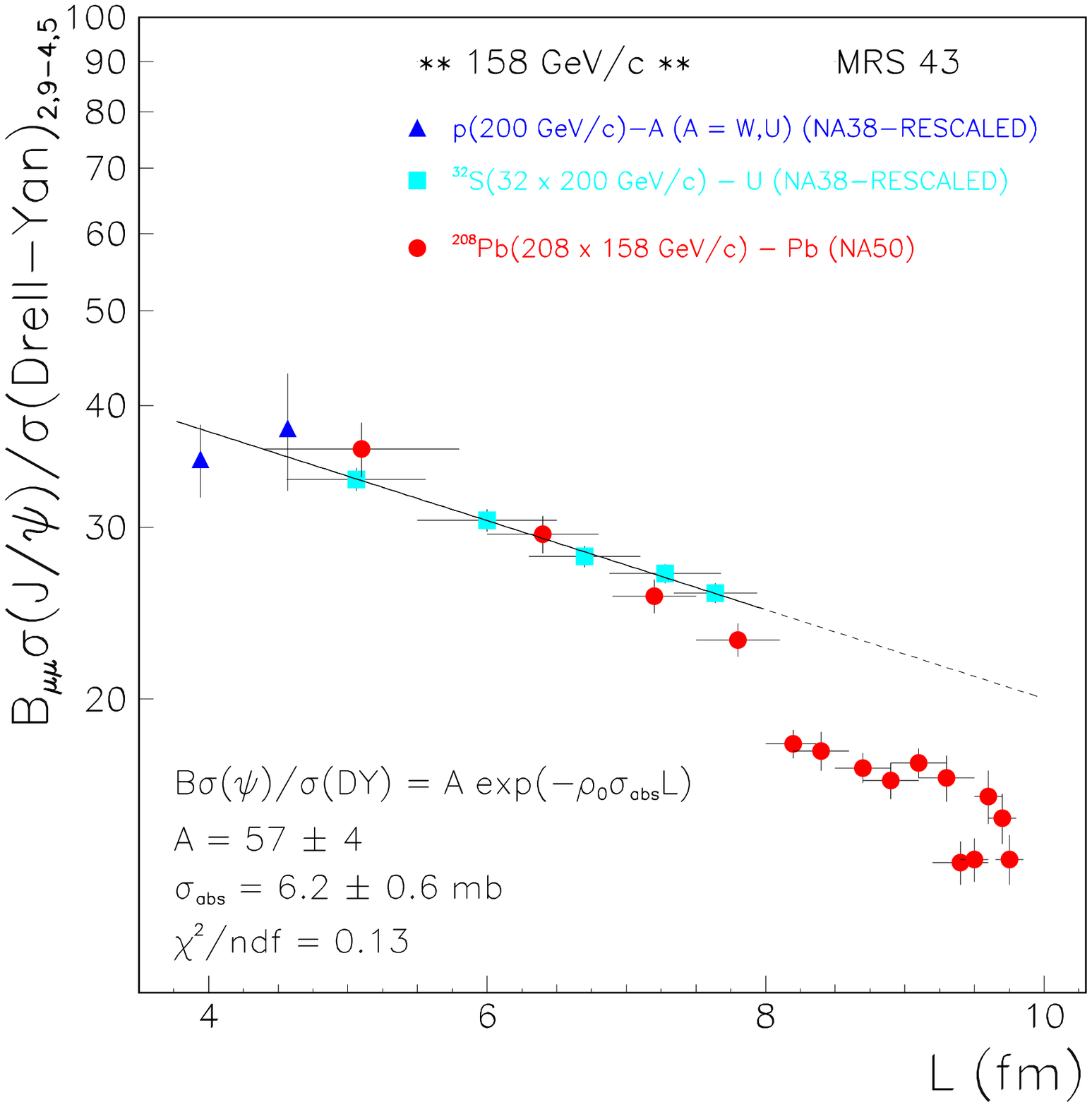, width=11cm}
\end{center}
 \caption{ }
Anomalous $J/\Psi$ suppression as observed by the NA50
collaboration \cite{na50_2}. For more details see text.
 \label{fig:na50}
\end{figure}

The NA50 collaboration reported anomalous $J/\Psi$ suppression in Pb+Pb
collisions at SPS energy at the Quark Matter '96 conference
\cite{na50_1}.  In 1996 the collaboration increased their data sample on
$J/\Psi$ by about a factor of 5, and modified their apparatus to cover a
much larger range in centrality. The preliminary results reported at
this conference are shown in Fig. \ref{fig:na50}. Shown here is the
ratio of $J/\Psi$ to Drell-Yan production versus the geometrical mean
path length L of the $J/\Psi$ or $\bar c c$ state traversing the target
and projectile matter. The quantity L is, apart from very central
collisions where L saturates, a fairly good measure of centrality but,
more importantly, can be used to compare various systems ranging from
p+A to Pb+Pb collisions. 

Prior to data on Pb+Pb collisions the ratio of
$J/\Psi$ to Drell-Yan was found to decrease exponentially with L,
implying absorption of the $J/\Psi$ or its precursory $\bar c c$ state
in the nuclear material with an absorption cross section of 6.2 $\pm
0.6$ mb. The new data shown in Fig.  \ref{fig:na50} confirm the findings
reported at Quark Matter '96 but now also demonstrate that, for
peripheral Pb+Pb collisions i.e. those with L less than about 8 fm, the
ratio of $J/\Psi$ to Drell-Yan follows approximately (actually the slope
seems somewhat steeper) the same exponential behavior also called
'normal nuclear absorption'. In fact, the data exhibit a rather strong
additional suppression setting in at around L = 8 fm. How rapid the
onset of anomalous absorption is needs to be studied further. However,
it is clear from the correlation between transverse energy and impact
parameter or L that fluctuations in transverse energy lead to
uncertainties of L (even for very small transverse energy bins) of the
order of 1 fm and, hence, no "discontinuity" can be expected in the data
over L ranges less than that.

Nevertheless, the most exciting interpretation of the anomalous
absorption is that, in the center of the hot and dense fireball formed
in the collision, the initially hadronic matter is converted into bubbles
of quark-gluon plasma where bound charmonium states cannot
survive. Certainly the observations are consistent with this scenario
(see, e.g., the talk by Kharzeev \cite{kharzeev} at this
conference). Alternative explanations have focussed on a scheme in which
hadronic comovers break up the $J/\Psi$ (see, e.g., the talk by Vogt
\cite{vogt} at this conference). However, within the framework of such
models it turns out rather difficult to explain the complete absorption
curve. Furthermore, for the absorption picture to work the comover
density must be unrealistically high (on the order of 1/fm$^3$).

\section{Summary and Outlook}
Over the past year the field has seen significant advances. There is now
strong evidence from many experiments that the initially very hot and
dense fireball formed in ultra-relativistic nuclear collisions expands
with a common flow velocity prior to freeze-out. This is based on the
analysis of single particle spectra as well as of two particle correlations
which we could not cover here for reasons of space. Other collective
features like directed flow and elliptic flow have now been also
demonstrated at SPS energy. Both at AGS and SPS energy thermal
freeze-out occurs at a temperature close to 120 MeV.  Chemical
freeze-out seems to occur at AGS energy at nearly the same temperature,
while at SPS energy the corresponding temperature is about 170 MeV:
apparently chemical freeze-out occurs close to where one expects the
phase boundary based on the predictions of lattice QCD and simple bag
models. New results from the CERES collaboration have consolidated the
picture on an enhanced low mass electron pair continuum, although a real
distinction between the various explanations put forward for the nature
of the enhancement will probably have to await new data with much
improved resolution and statistics. The $J/\Psi$ suppression observed by
the Na50 collaboration has not found any convincing explanation in terms
of conventional scenarios. Clearly an interesting ``picture'' of
ultra-relativistic nucleus-nucleus collisions is beginning to
emerge. Based on this we look into the future with 
great enthusiasm.

\end{document}